\documentclass[12pt,preprint]{aastex}
\newcommand\be{\begin{equation}}
\newcommand\ee{\end{equation}}
\newcommand\bea{\begin{eqnarray}}
\newcommand\eea{\end{eqnarray}}
\begin{document}

\title{Observational Constraints on Cardassian Expansion}

\author{S.Sen}
\affil{Centro de Astronomia e Astrofisica da Universidade de Lisboa (CAAUL),
 Departamento de F\'\i sica da FCUL, Campo Grande, 1749-016 Lisboa, Portugal.}
 \email{somasri@cosmo.fis.fc.ul.pt}
\and 
\author{A.A.Sen}
\affil{Centro Multidisciplinar de Astrof\'{\i}sica,
 Departamento de F\'\i sica, Instituto Superior T\'ecnico \\
 Av. Rovisco Pais 1, 1049-001 Lisboa, Portugal}
 \email{anjan@x9.ist.utl.pt}

\begin{abstract}
In this work, we have studied the observational constraints on the Cardassian Model
for the dark energy. We have compared the model with existing Supernova data. The dependence 
of the locations of the Cosmic Microwave Background Radiation (CMBR) peaks on the parameters
of the model have also been studied. We find, in particular, that observational data arising from 
Archeops for the location of the first peak, BOOMERANG for the location of the third peak, 
together with the Supernova data, constrain significantly the parameter space.
\end{abstract}

\vskip 5mm
\section{Introduction}
It is remarkable that number of current observations indicate that we are 
living in a spatially flat, low matter density universe  which is currently 
undergoing an accelerating expansion \citep{cmb1,cmb2,cmb3,super1,super2,super3,super4}. The most simple
 explanation of this current cosmological state of the universe requires two 
 dark components: one is in the form of non-relativistic dust (``dark matter'') 
 with vanishing pressure contributing one-third of the total energy density 
 of the universe and clustering gravitationally at small scales while the 
 second one is a smoothly distributed component having large negative 
 pressure (``dark energy'') and contributing around two-third of the total 
 energy density of the universe. Although the simplest candidate for this 
 dark energy is the vacuum energy or the cosmological constant $(\Lambda$), 
 alternative scenarios where the acceleration is driven by dynamical scalar 
 field  both minimally \citep{cald,peeb,
 fer,cope,stein,zlat,wett,rat,bar,sah,sen1,bento1} and non-minimally
 \citep{bertl,bert,uzan,amend1,gas1,sen2,sen3,sen4} coupled with gravity 
 called ``quintessence'' have been widely investigated 
 in recent years \footnote{See also \citep{ban1,ban2,soma,amend2,chib,gas2,riaz,toc}}. 

As none of the two components (dark matter and dark energy) has laboratory 
evidence both directly or indirectly, one has to invoke untested  physics 
twice to explain the current observations. That is why people in recent times 
have proposed interesting scenarios where one describes both dark matter and 
dark energy in a unified way through a single fluid component in the 
Einstein's equation. Chaplygin gas model is one such interesting possibility 
which has attracted lot of attentions in recent times
\citep{chap1,chap2,chap4,chap3}. 
Padmanabhan and Roy Choudhury  have also proposed an interesting unified 
description based on a  rolling tachyon arising in string theory \citep{paddy}.

Recently Freese \& Lewis\citep{freese} proposed another interesting alternative to quintessence scenario 
where the recent acceleration of the flat universe is driven solely by the  
matter, instead of using any cosmological constant or vacuum energy term. 
Since pure matter or radiation cannot alone take into account the 
recent acceleration in the flat universe, 
this goal is 
accomplished by modifying the Friedman equation with an empirical 
additional term named {\em Cardassian term}.
\be
H^2=A~\rho+B~\rho^n,
\ee
where $A={8\pi G\over{3}}$ and $B$ and $n$ are constants and are the 
parameters of the model. Here the energy density $(\rho)$ contains only matter 
$(\rho_m)$ and 
radiation $(\rho_r)$, i.e, $\rho=\rho_m+\rho_r$. Since at present 
$\rho_m>>\rho_r$, $\rho$ can be considered consisting of $\rho_m$ only. 
The new term, dominates only recently at redshift $\sim 1$. 
To provide the required acceleration of the universe 
as the outcome of the dominance of this term, $n$ should be $<2/3$.\\

There are several interpretations for the origin of this new ``cardassian term'' 
appearing in the Einstein's equation (1). As described in \citep{freese} and also 
in \citep{chung}, this term can appear as a consequence of embedding our observable 
universe  as a $3+1$ dimensional brane in extra dimension. Although recently it has 
been argued \citep{cline} that cardassian  model based on this 
higher dimensional interpretation violates the weak energy condition for the bulk 
stress energy for $n<2/3$ which is necessary for accelerating universe in late times. 
This extra term may also arise due to the matter self-interactions that contributes a 
negative pressure, through a long-range confining force which may be of gravitational 
origin or may be a fifth force \citep{gond}.
Also denoting the second term as $\rho_{x}$ one can recast equation (1) as,  
\be
H^2=A~(\rho+\rho_x),
\ee 
where one can consider $\rho_{x}$ as the dark energy with equation of state
\be
w_{x} = (n-1) + {n\over{3}}{\rho_{r0}a^{-4}\over{\rho_{r0}a^{-4}+\rho_{m0}a^{-3}}}
\ee
where $\rho_{r0}$ and $\rho_{mo}$ are the present energy density for radiation and 
matter and we have assumed that present scale factor $a_{0}=1$. At late times, 
when $ \rho_{r} << \rho_{m}$, one can approximate the above equation of state as 
\be
w_{x}^{late} = (n-1) + {n\over{3}}{\rho_{r0}\over{\rho_{m0}}}(1+z)
\ee
which is very slowly varying function with redshift. This is the conventional first 
order expansion to the equation of state for the dark energy and has been widely used 
in literature\citep{omega,ast,lind}. Also as $\rho_{r0}<<\rho_{m0}$, at late times, 
$w_{x}^{late}$ is almost constant
and it is identical  to a dark energy component with a constant equation of state\citep{av}.
But in early times, as one can not ignore the radiation component, one has to take the
general equation of state $w_{x}$  which is not constant. This is important when one considers
the constraints on the model from the CMBR observations. 

In this work, we shall consider the observational constraints on the different 
parameters of this cardassian model. We shall consider the constraints arising from 
the positions of the peaks of the CMBR as well as those arising from Supernova observations.

\section{Cardassian Model}
One can cast equation (1) in the following way  
\be
H^2=A~\rho[1+(\frac{\rho}{\rho_{car}})^{n-1}]
\ee
where $\rho_{car}$ is the energy density at which the two terms are equal. 
Once the energy density $\rho$ drops below $\rho_{car}$ the universe 
starts accelerating. $\rho_{car}$ is given by   
\bea
\rho_{car}&=&\left(\frac{A}{B}\right)^{\frac{1}{n-1}}\nonumber\\
&=&\rho_{m0}(1+z_{car})^{3}\{1+\frac{\Omega_{r0}}{\Omega_{m0}}(1+z_{car})\}
\eea
where $z_{car}$ is the redshift at which the second term, in equation (1), 
starts dominating over the first term. The model 
has two main parameters $B$ (or $\rho_{car}$ or $z_{car}$) and $n$.\\

To fulfill the requirement of the CMBR observation of a flat universe 
one can modify the critical energy density $\rho_c$ so that 
the matter can be sufficient to provide a flat geometry. Evaluating equation (1) today,
\be
H_0^2=A~(\rho_{m0}+\rho_{r0})\{1+\frac{B}{A}~(\rho_{m0}+\rho_{r0})^{n-1}\}.
\ee
In the new picture, $\Omega_0$ is defined as $\frac{\rho_{m0}+\rho_{r0}}
{\rho_c}$ so that matter alone makes the geometry flat. Here the 
expression for the critical energy density $\rho_c$ has been changed
 from its usual one $\rho_{ca}(=\frac{3H_0^2}{8\pi G})$, as,  
\be
\rho_c=\rho_{ca}\times F(n,z_{car})
\ee
where
\bea
F(n,z_{car})&=&[1+\frac{B}{A}\rho_{m0}^{n-1}(1+\frac{\Omega_{r0}}{\Omega_{m0}})^{n-1} ]^{-1}\nonumber\\
&=&[1+(1+z_{car})^{3(1-n)}\{1+\frac{\Omega_{r0}}{\Omega_{m0}}(1+z_{car})\}^{1-n}(1+
\frac{\Omega_{r0}}{\Omega_{m0}})^{n-1} ]^{-1}
\eea
$\Omega_{m0}$ and $\Omega_{r0}$ are two parameters  defined as 
$\Omega_{m0}=\frac{\rho_{mo}}{\rho_{ca}}$ and $\Omega_{r0}=\frac{\rho_{ro}}{\rho_{ca}}$ 
respectively.

So the new critical density $\rho_c$ is expressed as a function of two 
parameters $n$ and $B$ or $z_{car}$. This is similar to the expression given 
in \citep{freese}.
As mentioned there, a point to note here is that the new 
critical density $\rho_c$ is now the fraction of the original critical 
density $\rho_{ca}$ which 
has a standard value $1.88\times 10^{-29}~h_0^2~gm/cm^3$. Hence the new 
critical density can be much lower 
than the standard estimate. And also keeping in mind that we consider a 
flat geometry $(\Omega_0=1)$, we have today's energy density 
$\rho_{m0}+\rho_{r0}=\rho_c$
i.e,
\be
\Omega_{m0}+\Omega_{r0}=\frac{\rho_c}{\rho_{ca}}=~F
\ee
In figure 1, we present the different combinations of $n$ 
and $z_{car}$ for certain values of $F$.However, gravitational cluster\citep{abha} and other data suggest\citep{turner} the total matter density to be 30\% of the usual critical density i.e, $\rho_c=.3~\rho_{ca}$. This sets a preferred value $.3$ for $F$. 

\begin{figure}
\plotone{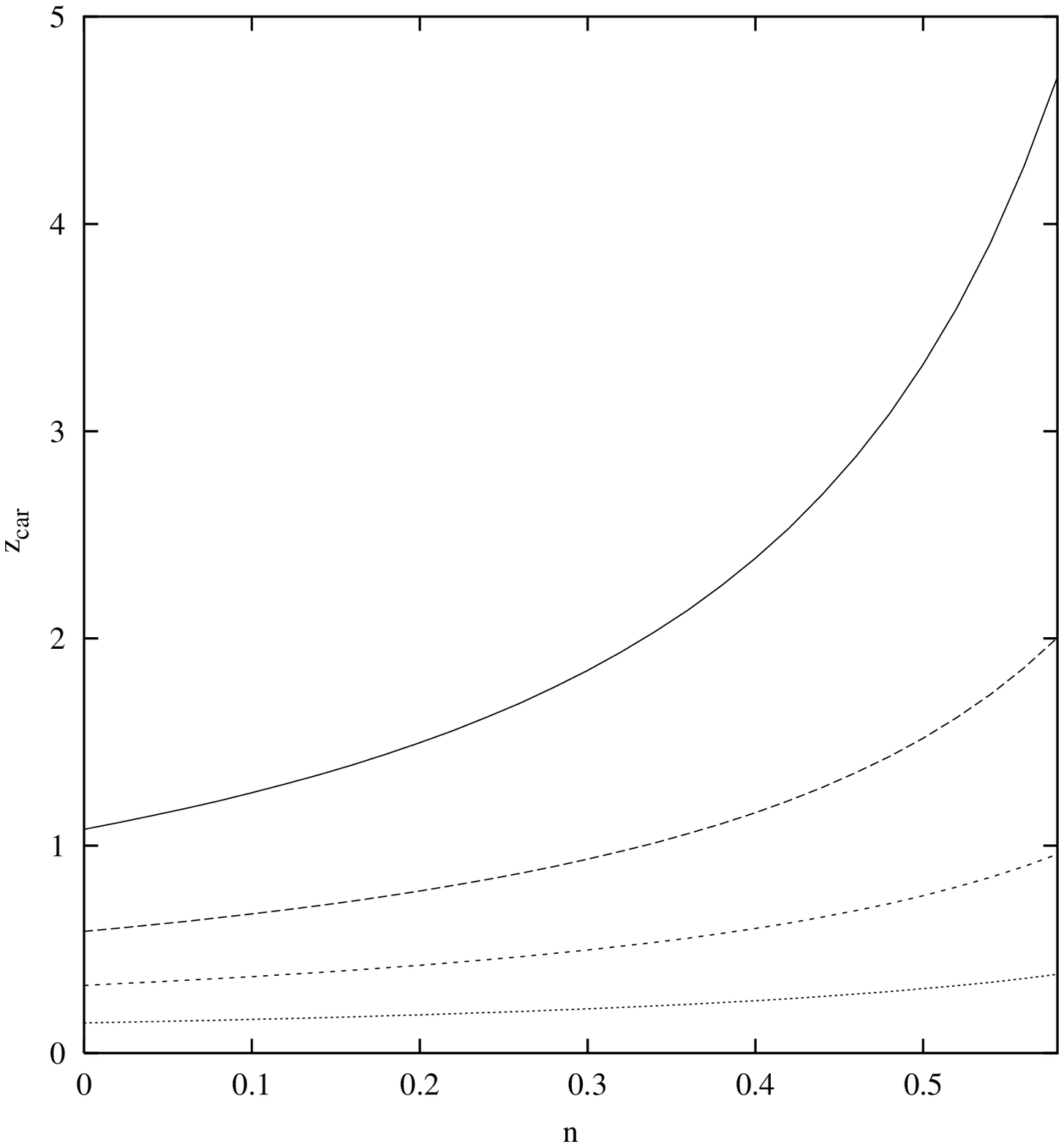}
\caption{Contours of F in ($n, z_{car}$)  plane. Contours corresponds
to value of F (0.1,0.2,0.3,0.4) from top to bottom.\label{fig1}}
\end{figure}

Now substituting the evolution of matter and radiation, we write equation (1) as the following,
\be
H^2=A~[\rho_{m0}a^{-3}(1+\frac{\Omega_{r0}}{\Omega_{m0}}a^{-1})+\frac{B}{A}
~\rho_{m0}^n~a^{-3n}(1+\frac{\Omega_{r0}}{\Omega_{m0}}a^{-1})^n]
\ee
From equation (9) it is very straight forward to express B in terms of 
$\Omega_{r0}$ and $\Omega_{m0}$ 

\be
\frac{B}{A}\rho_{m0}^{n-1}=(\frac{1-\Omega_{r0}-\Omega_{m0}}{\Omega_{m0}})(1+\frac{\Omega_{r0}}{\Omega_{m0}})^{-n}.
\ee
Substituting this expression in equation (11), one can finally 
recast equation (1) in the following fashion 
\be
H^2=\Omega_{m0}H^2_0 a^{-4}~\left[(a+\frac{\Omega_{r0}}{\Omega_{m0}})+a^{-4n+4}
\left(\frac{1-\Omega_{r0}-\Omega_{m0}}{\Omega_{m0}}\right)
\left(\frac{a+\frac{\Omega_{r0}}{\Omega_{m0}}}{1+\frac{\Omega_{r0}}{\Omega_{m0}}}\right)^n\right]
\ee
In equation (13) $H^2$ is expressed
in terms of the two model parameters $n$ and $\Omega_{m0}$ which, in turn, is 
related to the other form of this parameter ($F$ or $z_{car}$) through equation (9) and (10). 
We are now in a position to constrain these 
two parameters with different observations. 
Once we constrain these two parameters from the 
observation we can evaluate the corresponding value of $F$ or $z_{car}$ from eqn (10) and figure
\ref{fig1}.
There is an important point to note is that the value $n=0$ corresponds to a $\Lambda$CDM model
($\Omega_{r}$ at present is negligible). This will be crucial when we shall talk later about the allowed
region of the parameter space.

The prior assumptions in our subsequent calculations are as follows:  
scale factor at present $a_{0} = 1$, scale factor at last scattering 
$a_{ls} = 1100^{-1}$, $h = 0.65$, density parameter for radiation and baryons
at present $\Omega_{r0} = 9.89 \times 10^{-5}$, 
$\Omega_{b0} = 0.05$, 
and spectral index for the initial energy density perturbations, 
$n = 1$.

\section{Fitting with Supernova data}
First we check the consistency of the model with SNIa observations.
The data from different supernova observation is  related to a quantity called luminosity 
distance 
$d_l$ defined by
\be
d_l=(1+z)r_1
\ee
for a source at $r=r_1$ at $t=t_1$. But basically the luminosity distance 
in logarithmic units is what is observed by the astronomers.
\be
m_B(z)={\cal{M}}+5\log_{10}(D_l)
\ee
where ${\cal{M}}\equiv M-5\log_{10}H_0+25$ and $D_l=H_0 d_l$ is the 
dimensionless luminosity distance. To measure ${\cal{M}}$, one can 
show that for nearby sources (in low redshift limit) the above equation 
can be approximated as    
\be
m_B(z)={\cal{M}}+5\log_{10}(z).
\ee
The low redshift supernovae measurements can be used to calculate 
${\cal{M}}$. Using equation (14), we estimate the magnitudes of the 
supernovae at different redshifts from
\be
m_B-{\cal{M}}=5\log_{10}\{\frac{(1+z)}{\sqrt
{\Omega_{m0}}}\int_0^z\frac{dz'}{(1+z)^2 X(z)}\}
\ee
where 
$$
X(z)=\left[(\frac{1}{1+z}+\frac{\Omega_{r0}}{\Omega_{m0}})+(1+z)^{4n-4}
\left(\frac{1-\Omega_{r0}-\Omega_{m0}}{\Omega_{m0}}\right)\left(\frac{\frac{1}{1+z}+
\frac{\Omega_{r0}}{\Omega_{m0}}}{1+\frac{\Omega_{r0}}{\Omega_{m0}}}\right)^n\right]^{\frac{1}{2}}
$$

\begin{figure}
\plotone{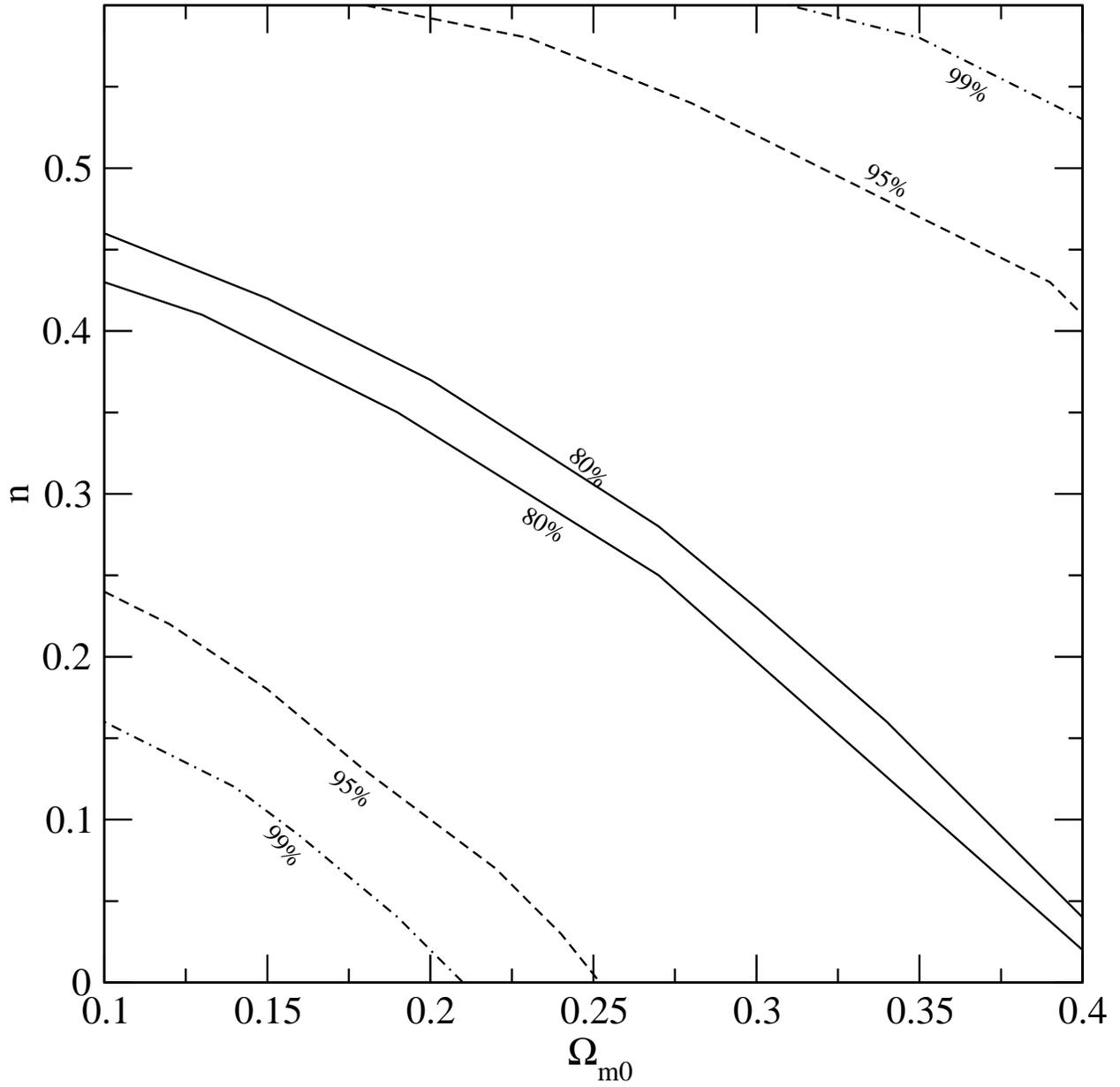}
\caption{The best fit contours for the Supernova data
at different confidence levels.\label{fig2}}
\end{figure}

Using the above the relation to estimate $m_B$ at different redshifts 
and the observed values of the effective magnitude $m_{B,i}^{eff}$ 
and the same 
standard errors $\sigma_{z,i}$ and $\sigma_{m_{B,i}^{eff}}$ for a given redshift 
as listed by Perlmutter et al from SCP\citep{super1}, we compute $\chi^2$ as
\be
\chi^2=\sum_{i=1}^{54}\frac{[m_{B,i}^{eff}-m_B(z_i)]^2}{\sigma_{z,i}^2+{\sigma^2_{m_{B,i}^{eff}}}} 
\ee
We consider the data set of 54 supernovae comprising of 38 high redshift 
supernovae from Supernova Cosmology Project together with 16 low 
redshift supernovae from the Calan-Tololo project, as used by 
\citep{super1} in their primary fit C (for details of 
excluded data points see \citep{super1}). In figure \ref{fig2} we   
present the permitted parameter space by the supernova constrains 
at different confidence level.
 We observe that 
the model best fits the current supernova data at 80\% confidence level.

\section{Constraints from CMBR}
Our second tool for constraining the parameters is CMBR anisotropy spectrum. 
The CMBR peaks arise from oscillation of the primeval plasma just before 
the universe becomes transparent. The oscillation of the tightly bound 
photon-baryon fluid is a result of the balance between the gravitational 
interaction and photon pressure and this oscillations gives rise to the 
peaks and troughs in the temperature anisotropic spectrum. In an ideal   
photon-baryon fluid model, there is an analytic relation for the location 
of the m-th peak\citep{hu}:
\be
l_m=m~l_A
\ee 
where $l_A$ is the acoustic scale which depends on both pre and post 
recombination physics and also on the geometry of the universe. This has an 
analytical expression given by  $\frac{\pi D}{s_{ls}}$ where $D$ is 
the angular diameter distance to the last scattering and $s_{ls}$ 
is the sound horizon at the last scattering. In terms of the conformal time 
$\tau$, $l_A$ is given by\citep{doran2,doran4},  
\be  
l_A=\pi\frac{\tau_0-\tau_{ls}}{\bar{c_s}\tau_{ls}}.
\ee 
where $\tau_0$ and $\tau_{ls}$ are the conformal time today and at 
last scattering and $\bar{c_s}$ is the average sound speed before 
last scattering. $\bar{c_s}$ is a constant for a particular 
$\frac{\rho_b}{\rho_r}$. We take it as 0.52 as others\citep{doran2}. \\

Now, to find $l_A$, we write equation (13)
in terms of conformal time 
\be
(\frac{da}{d\tau})^2=
\Omega_{m0}H^2_0~\left[(a+\frac{\Omega_{r0}}{\Omega_{m0}})+a^{-4n+4}
\left(\frac{1-\Omega_{r0}-\Omega_{m0}}{\Omega_{m0}}\right)\left(\frac{a+
\frac{\Omega_{r0}}{\Omega_{m0}}}{1+\frac{\Omega_{r0}}{\Omega_{m0}}}\right)^n\right]
\ee
wherefrom it is quite easy to find
\be 
\tau_{ls}=\int_0^{\tau_{ls}} d\tau=\frac{1}{\Omega_{m0}^{1/2}H_0}\int_0^{a_{ls}}\frac{da}{X(a)} 
\ee
and
\be 
\tau_{0}=\int_0^{\tau_{0}} d\tau=\frac{1}{\Omega_{m0}^{1/2}H_0}\int_0^1 \frac{da}{X(a)} 
\ee
where $X(a)=\sqrt{(a+\frac{\Omega_{r0}}{\Omega_{m0}})+a^{-4n+4}
\left(\frac{1-\Omega_{r0}-\Omega_{m0}}{\Omega_{m0}}\right)
\left(\frac{a+\frac{\Omega_{r0}}{\Omega_{m0}}}{1+\frac{\Omega_{r0}}{\Omega_{m0}}}\right)^n}$.

Substituting the above expression in equation (20), we have the analytical 
expression for $l_A$ in case of this model
\be
l_A=\frac{\pi}{\bar{c_s}}\left[\frac{\int_0^1 \frac{da}{X(a)}}{\int_0^{a_{ls}} \frac{da}{X(a)}}-1\right]
\ee
where $a_{ls}=1100^{-1}$. Thus we can find the positions of the peaks in the 
CMBR spectrum from equation (19). $l_A$ and consequently the positions 
of the CMBR peaks here
depends only $n$ and $\Omega_{m0}$. So 
once we calculate the positions of the peaks we can constrain the 
parameters $n$ and $\Omega_{m0}$ by comparing the results from 
different observations.

Now the simple relation in (19) is modified by driving and dissipative 
effects which introduces a phase shift to the oscillation such that\citep{hu} 
\be
l_m\equiv l_A(m-\phi_m)
\ee
The phase shift of the peaks $\phi_m$ is predominantly determined by the 
pre-recombination physics and is independent of the geometry of the 
universe. It depends on parameters like, $\Omega_b h^2$, $n$, and $r_{ls}$, 
the ratio of the energy density of radiation to matter at last scattering. 
For $n=1$ and $\Omega_b h^2=0.02$ and $r_{ls}=\frac{\rho_r(z_{ls})}
{\rho_m(z_{ls})}$, $z_{ls}\sim 1100$ at last scattering,     
the phase shift for the first peak\citep{hu}  
\be 
\phi_1= 0.267(\frac{r_{ls}}{0.3})^{.1}. 
\ee
Now substituting for $r_{ls}$ at $z=1100$, we have
\be
\phi_1= 0.267(\frac{1100\Omega_{r0}}{0.3\Omega_{m0}})^{.1}. 
\ee
Hence the position of the first peak $l_1$ is
\be
l_1=l_A(1-\phi_1)
\ee
Using equation (22)-(24) and (27) and (28) one can calculate $l_1$ 
as a function of $n$ and $\Omega_{m0}$. The observational bounds on 
$l_1$ as predicted by BOOMERANG\citep{boom} and more recently by Archeops 
\citep{arc} are $l_1=221\pm14$ and $l_1=220\pm6$ respectively.
As the bound coming from the Archeops data is more stringent, we shall take 
this bound for first peak to constrain our parameters.

\begin{figure}
\plotone{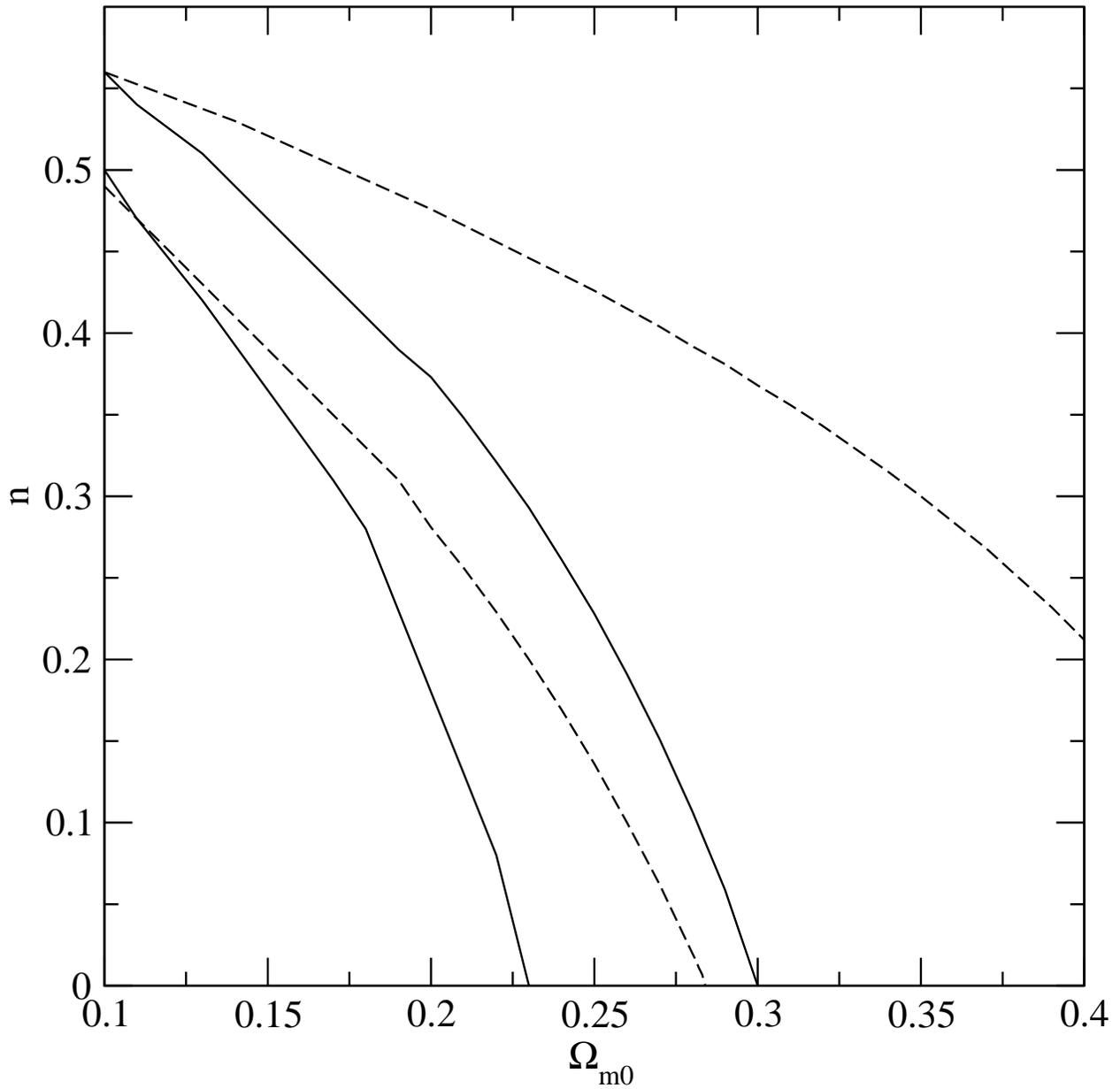}
\caption{Contours in the ($n,\Omega_{m0}$)  plane arising from Archeops 
constraints on $l_{1}$(dashed contour) and BOOMERANG constraints on $l_{3}$(full contour), and Supernova observations.
 The allowed region of the model parameters lies in the intersection between these
 regions.\label{fig3}}
\end{figure}

The relative shift of the second peak is a very sensitive 
quantity and depends on many parameters. Hence it is very 
difficult to derive any constraint from the second peak. 
So we disregard the second one.  
As far the third peak 
is concerned Doran et al\citep{doran3} 
have shown it to be insensitive to different cosmological 
parameters. They estimated $\phi_3$ to be $0.341$. Hence
\be
l_3=l_A(3-\phi_3)=l_A(3-0.341)=2.659~l_A
\ee
With this expression one can also calculate $l_3$ for different 
values of $n$ and $\Omega_{m0}$. The observational bound on $l_3$ as suggested by 
BOOMERANG : $l_3=845^{+12}_{-25}$. In figure \ref{fig3} we have shown the constraints on the parameter space
that are obtained from the observational bounds on the location of the first(dashed contour)
 and third(full contour) CMBR peaks. Hence, from the CMBR point of view the allowed region of the
 model parameters lies in the intersection between these two contours.

\section{Conclusion}

\begin{figure}
\plotone{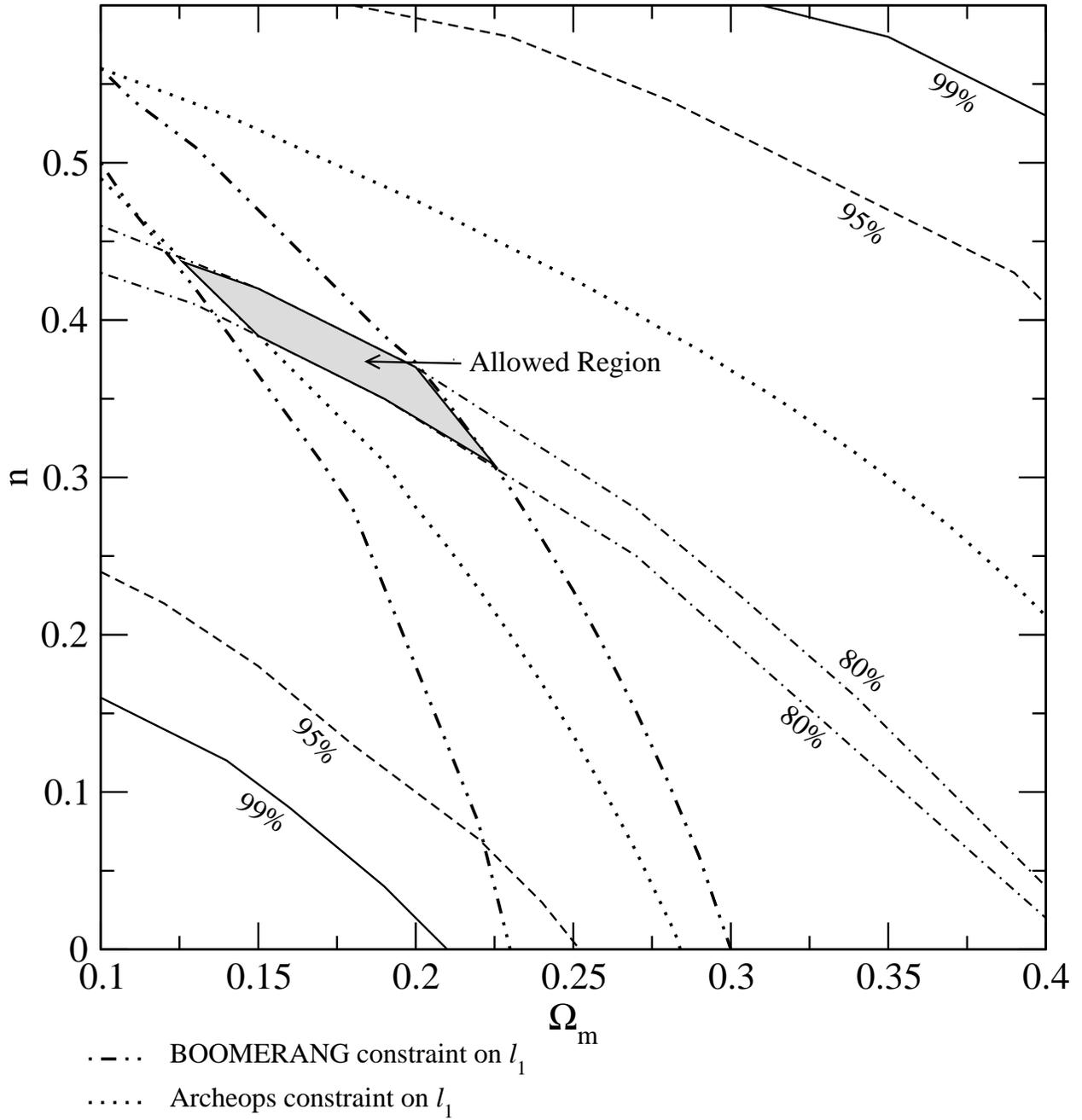}
\caption{The allowed region in (n,$\Omega_{m}$) plane 
arising from the joint analysis.\label{fig4}}
\end{figure}

In conclusion, we have shown that the locations of the CMBR peaks, as determined
via Archeops and BOOMERANG, as well as the present Supernova Ia data, constrain a sizable portion of the parameter space of the Cardassian model.
We observe that the model best fit with the Supernova data at 80\% confidence 
level. This, together with the allowed region from CMBR data restricts the parameters as
$0.31 \lesssim n \lesssim 0.44$ and $0.13 \lesssim \Omega_{m0} \lesssim 0.23$ (See figure \ref{fig4}). This clearly does not 
include the $n=0$ which is the corresponding $\Lambda$CDM case in this model. This is the most 
interesting result of this investigation. Also the joint analysis indicates a lower value of $\Omega_{m0}$.
This is consistent with that predicted by Zhu et.al \citep{zhu} by investigating the constraints on the
cardassian model from the recent measurements of the angular size of high-z compact radio sources. 
Also our bound on $\Omega_{m0}$ is consistent with that 
predicted by Melchiorri et.al 
\citep{mel} from a combined CMB+HST+SNIa+2dF analysis. Clearly with future high precision measurements 
of the 
MAP and PLANCK mission, we expect that the positions of the CMBR peaks will be determined
with higher accuracy. This, together with the upcoming data from future SNAP mission will
further constrain the parameter space of this model.
\vskip 0.4cm
\acknowledgments
The authors are grateful to O. Bertolami, M.C. Bento and P. Crawford
 for useful discussions. The authors also thank Zong-Hong Zhu to bring in our attention 
 one of their recent papers, and also for his comments.
The work of A.A.S. is fully 
financed by  Funda\c c\~ao para a Ci\^encia e a Tecnologia (Portugal)
under the grant POCTI/1999/FIS/36285. The work of S.S. is financed by 
 Funda\c c\~ao para a Ci\^encia e a Tecnologia (Portugal), through CAAUL.
\newpage

\clearpage
\clearpage
\clearpage
\clearpage
\end{document}